\documentclass[5p]{elsarticle}
\biboptions{sort&compress}
\usepackage{amsmath}
\usepackage{amssymb}
\usepackage{hyperref}
\def\WpWm{W^+W^-}
\def\sqrts{\sqrt{s}}
\def\qT{q_T}
\def\pTlp{p_{\perp,\ell^+}}
\def\mll{m_{\ell^+ \ell^-}}
\def\mTWW{m_{T,WW}}
\def\Delphi{\Delta \phi_{\ell^+ \ell^-}}
\newcommand{\mrm}{\mathrm}
\newcommand{\beq}{\begin{equation}}
\newcommand{\eeq}{\end{equation}}
\journal{Physics Letters B}
\begin{document}
\begin{frontmatter}

\title{
{%
\vspace{-1.0cm}
\small\hfill\parbox{64.0mm}
{\raggedleft%
CERN-PH-TH-2015-276, TTP15-041
}}\\[0.5cm]
QCD corrections to $W^+W^-$ production through gluon fusion}

\author[cern]{Fabrizio Caola}            
\ead{fabrizio.caola@cern.ch}

\author[kit]{Kirill Melnikov}            
\ead{kirill.melnikov@kit.edu}

\author[kit]{Raoul R\"ontsch}            
\ead{raoul.roentsch@kit.edu}

\author[kit]{Lorenzo Tancredi}             
\ead{lorenzo.tancredi@kit.edu}

\address[cern]{CERN Theory Division, CH-1211, Geneva 23, Switzerland}
\address[kit]{Institute for Theoretical Particle Physics, KIT, Karlsruhe, Germany}

\begin{abstract}
We compute the next-to-leading order (NLO) QCD corrections to the 
$gg \to \WpWm \to \l^+_1\nu_1 l^-_2 \bar\nu_2$ process, mediated by a massless quark loop, at the LHC.
This process first contributes to the  hadroproduction of $\WpWm$ at $\mathcal{O}(\alpha_s^2)$, 
but, nevertheless,  has a sizable impact on the total production rate.
We find that the NLO QCD corrections to the $gg \to W^+W^-$  process amount to ${\cal O}(50)$\%, 
and increase the NNLO QCD cross sections of $pp \to W^+W^-$ by approximately two percent, at both the 8 TeV and 13 TeV LHC. 
We also compute the NLO corrections to gluonic $\WpWm$ production within a fiducial volume used  by the ATLAS
collaboration in their 8 TeV measurement of the $W^+W^-$ production rate 
and find that the QCD corrections are significantly smaller than in the inclusive case.
  While the current 
experimental uncertainties  are still too large to make these differences relevant, the observed strong 
dependence of  perturbative 
corrections on kinematic cuts underscores that extrapolation from a fiducial measurement  to the total cross section 
is an extremely delicate matter, 
and calls for the direct comparison of fiducial volume measurements with corresponding theoretical computations.
\end{abstract}
\end{frontmatter}

The production of electroweak di-bosons, $pp \to VV$, 
is amongst the most important processes studied at the LHC.
The Higgs decay mode $H \to VV$ will be central to precision measurements of the Higgs 
quantum numbers and couplings during Run II~\cite{Aad:2014eva,Aad:2015xua,Aad:2015rwa,
Aad:2015tna,Khachatryan:2014jba,Khachatryan:2014kca,Khachatryan:2014iha,Khachatryan:2015mma}.
This requires extremely good control over the large $pp \to VV$ background, including in the Higgs 
off-shell region \cite{Kauer:2012hd},  which can  be exploited 
to constrain $HVV$ couplings~\cite{Azatov:2014jga} or the Higgs width~\cite{Caola:2013yja,Campbell:2013una,Campbell:2013wga}.
Additionally, di-boson production probes the nature of the electroweak interactions, 
allowing New Physics effects to be either discovered or constrained through studies of anomalous gauge couplings. 
Finally, di-boson production serves as an important testing ground for our understanding of QCD in a collider environment.

At leading order (LO), weak boson pair production $pp \to VV$ occurs only through the $q\bar{q}$ partonic channel.
The next-to-leading order (NLO) QCD corrections to this process have been studied 
extensively in the past~\cite{Mele:1990bq,Ohnemus:1990za,Ohnemus:1991kk,Frixione:1993yp,Ohnemus:1994,Dixon:1998py,Campbell:1999ah,Dixon:1999di,Campbell:2011bn}; 
recently the next-to-next-to-leading order (NNLO) QCD corrections have 
also been computed~\cite{Catani:2011qz,Grazzini:2013bna,Cascioli:2014yka,Gehrmann:2014fva,Grazzini:2015nwa,Grazzini:2015hta}.
At this order, the $gg$ partonic channel 
starts contributing~\cite{Dicus:1987dj,Glover:1988rg,Glover:1988fe,Binoth:2005,Binoth:2006mf,Binoth:2008} and, thanks to a relatively large gluon flux 
at the LHC, its contribution can be expected to be large. This is exactly what happens: 
the gluon fusion process contributes 60\% of the 
NNLO QCD corrections in $ZZ$ production, and 35\% of the  NNLO QCD corrections in $\WpWm$ production. 
Radiative corrections to the gluon fusion channel formally contribute at 
next-to-next-to-next-to-leading order (N$^3$LO) but  are expected to 
be significant~\cite{Bonvini:2013jha}.
Indeed, we showed recently that NLO QCD corrections to $gg \to ZZ$ increase its contribution 
to $pp \to ZZ$ by almost a factor  of two making them important for phenomenology of $ZZ$ production~\cite{Caola:2015psa}. 
Moreover, the magnitude of these  corrections exceeds the scale variation uncertainty of the  NNLO QCD result 
which is commonly used to estimate the residual uncertainty of the theory prediction. 
The aim of this Letter is to report the results of a similar 
calculation for $gg \to \WpWm$. 

Run I measurements of the $\WpWm$ cross section undertaken 
by both ATLAS~\cite{ATLAS:2012mec} and CMS~\cite{Chatrchyan:2011tz,Chatrchyan:2013oev}
showed a discrepancy at the level of ${\cal O}(2-2.5)$ standard deviations 
compared to the Standard Model (SM) prediction.
This deviation has been studied in the context of physics beyond the 
Standard Model (BSM)~\cite{Curtin:2012nn,Curtin:2013gta,Rolbiecki:2013fia,Jaiswal:2013xra,Curtin:2014zua,Kim:2014eva},
but there has also 
been a concerted effort from the theory community to understand the source of this discrepancy in terms of QCD effects.
This includes the calculation of the total $\WpWm$ cross 
section to NNLO in QCD~\cite{Gehrmann:2014fva},
as well as the examination of ambiguities caused  by an extrapolation  from the fiducial 
region to the total cross section, 
either in the context of parton showers~\cite{Monni:2014zra} or 
through resummations~\cite{Meade:2014fca,Jaiswal:2014yba,Jaiswal:2015vda}.
As a consequence of these efforts, the 
discrepancy seems to have been resolved without recourse to BSM effects, 
but these developments underlined  the importance of  comparing theoretical and 
experimental results for fiducial volume measurements, avoiding uncertainties 
related to the extrapolation.  Motivated by these considerations, we also study  
the NLO QCD corrections to  $gg \to \WpWm$  in the fiducial region defined by the ATLAS cuts.

\begin{figure}
\centering
\includegraphics[scale=0.2]{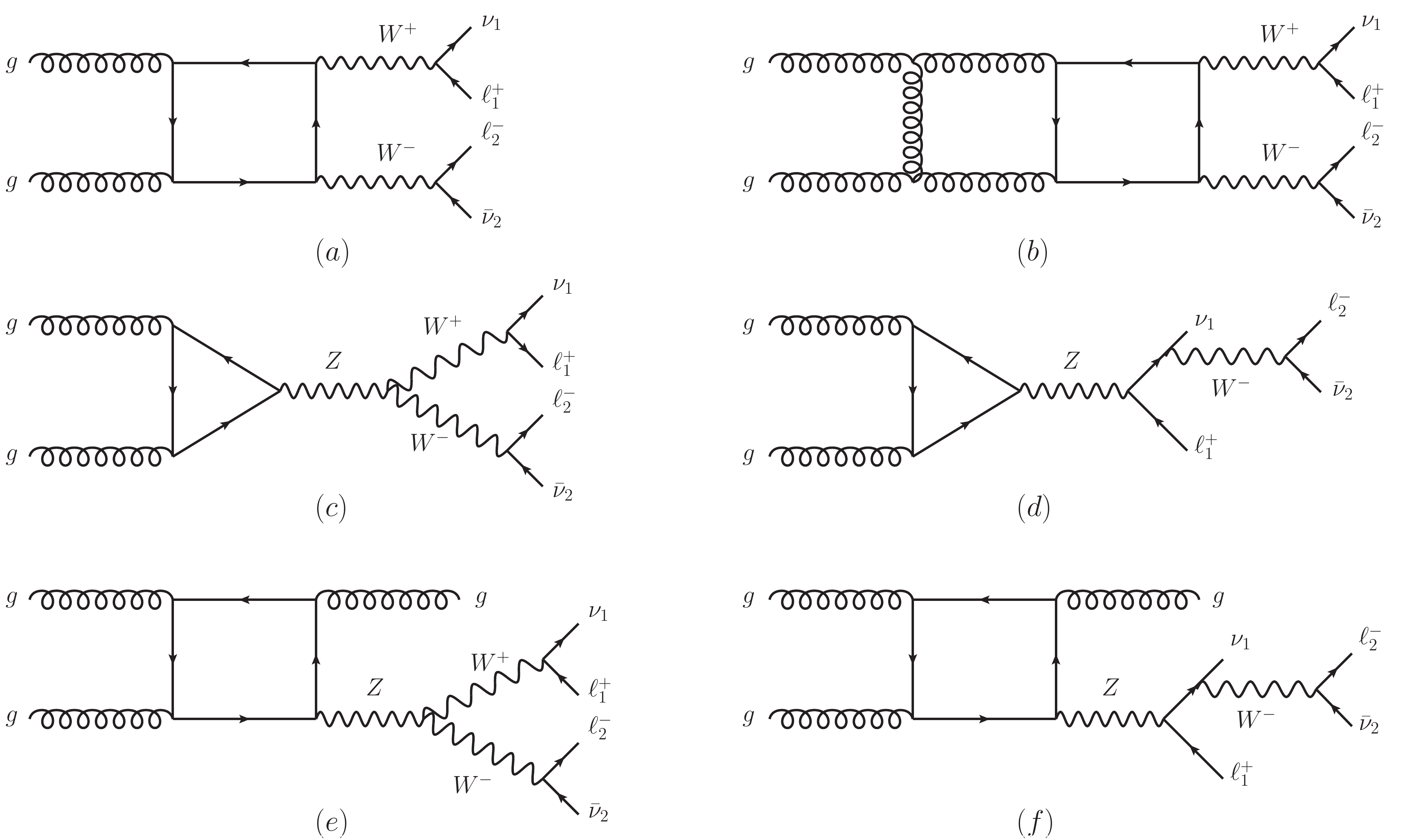} \\
\caption{\label{fig:ampl} Representative Feynman diagrams that contribute to gluon fusion 
process $gg \to \nu_1 \ell_1^+ \ell_2^- \bar{\nu_2}$ through NLO in perturbative QCD.   }
\end{figure}

We begin by summarizing the technical details of the calculation, and refer the reader to Ref.~\cite{Caola:2015psa} for a more
extensive discussion. 
In Fig.~\ref{fig:ampl} we present representative Feynman diagrams that are required for the calculation 
of the gluon fusion process $gg \to \WpWm$ through NLO in perturbative QCD. 
The required two-loop contributions to $gg \to \WpWm$ scattering amplitudes
have been calculated in Refs.~\cite{Caola:2015ila,vonManteuffel:2015msa}; we use the {\sf C++} code 
developed in Ref.~\cite{vonManteuffel:2015msa} in our computation. 
We calculate the relevant one-loop real-emission amplitudes $gg \to \WpWm + g$ 
using a combination of numerical~\cite{Ellis:2007br} and analytic~\cite{Badger:2008cm} unitarity methods.
The virtual and real-emission contributions are 
combined using both the $\qT$~\cite{Catani:2007vq} and the FKS~\cite{Frixione:1995ms} subtraction schemes, 
allowing for a check of the numerical stability of the final  results and the consistency of the 
implementation in a numerical program. 

Throughout the paper,  we consider leptonic decays of the $W$-bosons,  $gg \to \WpWm \to \nu_1 \ell_1^+ \ell_2^- \bar{\nu_2}$. 
We note that if the leptons are of the same flavor and off-shell contributions are allowed then, strictly 
speaking,   it is impossible to distinguish off-shell $\WpWm$ production from off-shell $ZZ \to \ell^+ \ell^- \nu \bar{\nu}$ production, 
so that both contributions need to be included. We do not consider this issue here and postpone its investigation to the near future.

We will now comment on the various contributions  to $gg \to W^+W^-$ amplitudes. 
In addition to typical box-type amplitudes shown  in Fig.~\ref{fig:ampl}(a), we need to consider 
amplitudes where gluons couple to 
$Z^*$ and/or $\gamma^*$ through  the quark loop,  see  Figs.~\ref{fig:ampl}(c),(d). 
However,  it was shown in Ref.~\cite{Campbell:2007ev} that the 
sum of these triangle diagrams  vanishes to all loop orders 
for on-shell colliding gluons. This implies that we  only have to consider 
$Z/ \gamma^*$-mediated amplitudes if an additional gluon is 
 radiated,  c.f. Figs.~\ref{fig:ampl}(e-f).
Note that we also include the singly-resonant amplitudes in our calculation, see Fig.~\ref{fig:ampl}(f).

The most important difference with respect to our previous work is the treatment of the massive quark loops.
In $ZZ$ production, it is possible to separate the contribution of bottom and top loops, 
if one neglects the contribution of vector-axial triangle diagrams which are suppressed by the top mass.
One can then consider gluon-initiated $ZZ$ production through loops of five  massless quark 
flavors. On the other hand, if $W$ bosons are radiated from the quark loop, 
such a separation is obviously not possible since $W$-bosons mediate transitions within 
a given generation and mix the contributions of bottom and top quarks.
Since we cannot compute two-loop diagrams with internal masses, for $gg \to W^+W^-$ amplitudes 
we neglect the contribution of the third generation entirely, 
and consider only massless quarks of the first two generations.
However, for real-emission amplitudes which involve a $Z/ \gamma^*$ 
boson attached to  the quark loop (Fig.~\ref{fig:ampl}(e) and (f)), we 
adopt our previous approach and also include massless bottom quarks in the loop. We expect that 
the accuracy of this approach is ${\cal O}(10\%)$; 
this estimate is based on the observation that the inclusion of the third generation 
in the computation of the $gg \to W^+W^-$ leading order cross section changes the result 
by approximately this amount~\cite{Campbell:2011cu,gg2VV}. We have checked that the ratio  of leading order 
cross sections 
$\sigma_{gg \to W^+W^-}^{3 \rm gen}/\sigma^{2 \rm gen}_{gg \to W^+W^-} \approx 1.1$ is practically  
independent of the collision energy and the 
kinematic cuts that we use later on to identify the fiducial volume cross section. This observation 
offers a simple way to account for the effect of the third generation: 
although   we  present  the numerical results 
below omitting the contribution of the third generation, it can be included in an approximate 
way by increasing all our results by ten percent. This is the best one can do as long as the NLO QCD corrections  
to $gg \to W^+W^-$ process, mediated by  a massive quark loop, remain unknown.

We now present the results for $gg \to \WpWm \to \nu_e e^+\mu^- \bar{\nu}_{\mu}$ cross sections.
To perform the computation, we take the masses of 
the $W$ and $Z$ bosons to be $m_W=80.398$ GeV and $m_Z=91.1876$ GeV, their  
widths to be  $\Gamma_W=2.1054$ GeV and $\Gamma_Z=2.4952$~GeV and the Fermi constant 
$G_F = 1.16639~{\rm GeV}^{-2}$.
We use  $\mu=\mu_0 = m_W$ as the central value  for the renormalization and factorization scale, 
and estimate the effect of the scale variation by calculating the cross section at  $\mu = 2 \mu_0$ 
and $\mu = \mu_0/2$.  We use LO and NLO NNPDF3.0 parton distribution functions~\cite{Ball:2014uwa}, accessed 
through LHAPDF6~\cite{Buckley:2014ana} and 
one- and two-loop running of the strong coupling, for our LO and NLO results, respectively.
We do not include the contribution from Higgs-mediated amplitudes. Unless stated otherwise,
$W$ bosons are produced on the mass shell.
At $\sqrts=8$ TeV, we find the inclusive cross sections at LO and NLO to be
\beq
\sigma^{\WpWm}_{gg,\mrm{LO}} = 20.9^{+6.8}_{-4.8}~\mrm{fb},\;\;\; \sigma^{\WpWm}_{gg,\mrm{NLO}} = 32.2^{+2.3}_{-3.1}~\mrm{fb},
\label{xsect8}
\eeq
where the superscript (subscript) refers to the value at $\mu=\mu_0/2$ ($\mu=2\mu_0$).
We note that the NLO QCD corrections increase the gluon fusion  
cross section by a factor of $1.24-1.80$,  
with an increase by a factor of 1.54 for the central scale choice. 
This is similar to what was found in Ref.~\cite{Caola:2015psa} 
for $gg \to ZZ$ production, when one takes into account the different choice made there for the central scale.

\begin{figure}
\centering
\includegraphics[scale=0.5]{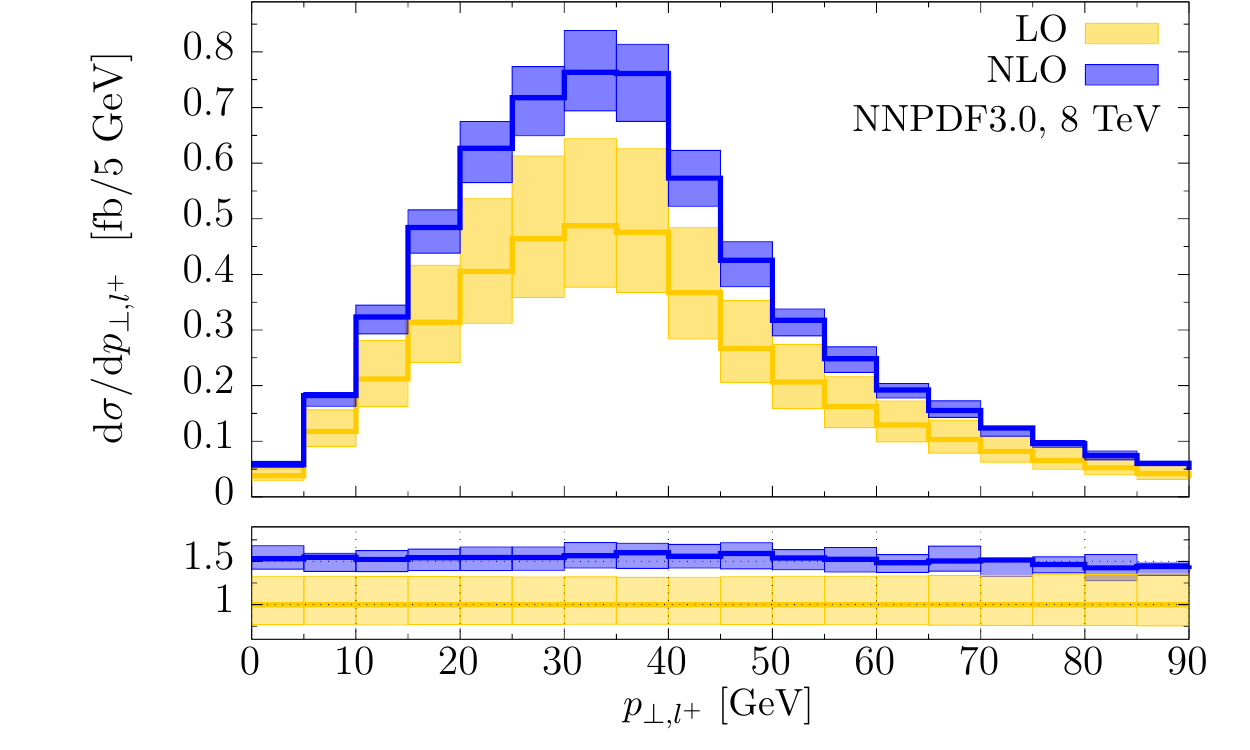}\\
\includegraphics[scale=0.5]{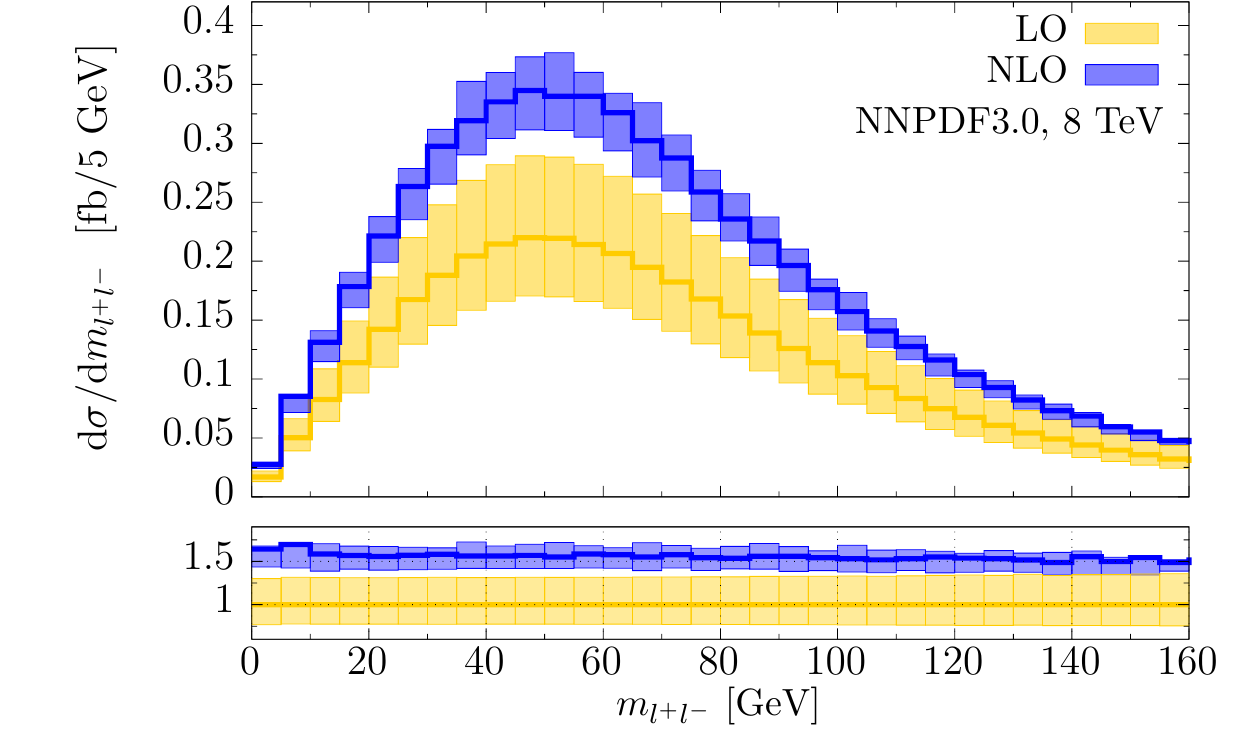}
\caption{\label{fig:distr1} The transverse momentum of the positron $\pTlp$ (upper plot) 
and the invariant mass of the dilepton system $\mll$ (lower plot) in 
$gg \to \WpWm \to  \nu_e e^+\mu^- \bar{\nu}_{\mu}$ process at the $\sqrt{s}=8$ TeV LHC. LO 
results are shown in yellow, NLO results are shown in blue. 
The central scale is $\mu=m_W$; the scale variation bands correspond to  scale variations by a factor of two in 
either direction. 
The lower panes show the ratios of the LO and NLO distributions at each scale to the LO distribution at the central scale.}
\end{figure}

In order to put these results into context, we would like to estimate their impact on the 
NNLO QCD prediction  for  the $pp \to W^+W^-$ process 
at $\sqrts = 8~$TeV
presented recently in Ref.~\cite{Gehrmann:2014fva}.
The results reported in  Ref.~\cite{Gehrmann:2014fva} were obtained for stable 
$W$-bosons; to compare them with our results, we have to multiply them by the branching fractions 
for $W$ decays into leptons. With the input parameters described above we find ${\rm Br}(W \to \ell \nu_l) = 0.108$,
in good agreement with experimental measurements. 
Then, taking the cross sections from  Ref.~\cite{Gehrmann:2014fva} at  $\mu = \mu_0$, we obtain
\beq
\sigma_{\mrm{NLO}} = 638.84~\mrm{fb};\;\;\; \sigma_{\mrm{NNLO}+gg,\rm LO} =697.97~\mrm{fb}.
\eeq
It is stated in Ref.~\cite{Gehrmann:2014fva} that about 35\% of the NNLO QCD corrections is due to 
 the gluon fusion channel;  this implies that the $gg \to W^+W^- \to 2\ell 2\nu$ cross section used in 
Ref.~\cite{Gehrmann:2014fva} is ${\cal O}(21)~{\rm fb}$ which compares well with our result 
in Eq.(\ref{xsect8}).  We now substitute the NLO QCD result  for the gluon fusion cross section 
instead of the LO one and obtain\footnote{We note that including contributions of the third generation would increase
this cross section by approximately 2 fb.}
\beq
\sigma_{\mrm{NNLO}+gg,\mrm{NLO}} \approx 710~\mrm{fb}.
\eeq
Therefore, inclusion of the NLO corrections to 
the gluon-initiated partonic channel  increases the total NNLO QCD cross section by about $2\%$ percent. 
This shift is comparable to the residual theoretical uncertainty on the NNLO QCD prediction 
for $pp \to W^+W^-$, which is quoted as ${\cal O}(2\%)$ in Ref.~\cite{Gehrmann:2014fva}.
We also note that the shift is much larger than the off-shell cross section for Higgs 
boson production $gg \to H^* \to W^+W^- \to 2\ell 2\nu$ which we estimate to be ${\cal O}(1)~{\rm fb}$
using the MCFM program~\cite{mcfm}.
\begin{figure}[t]
\centering
\includegraphics[scale=0.5]{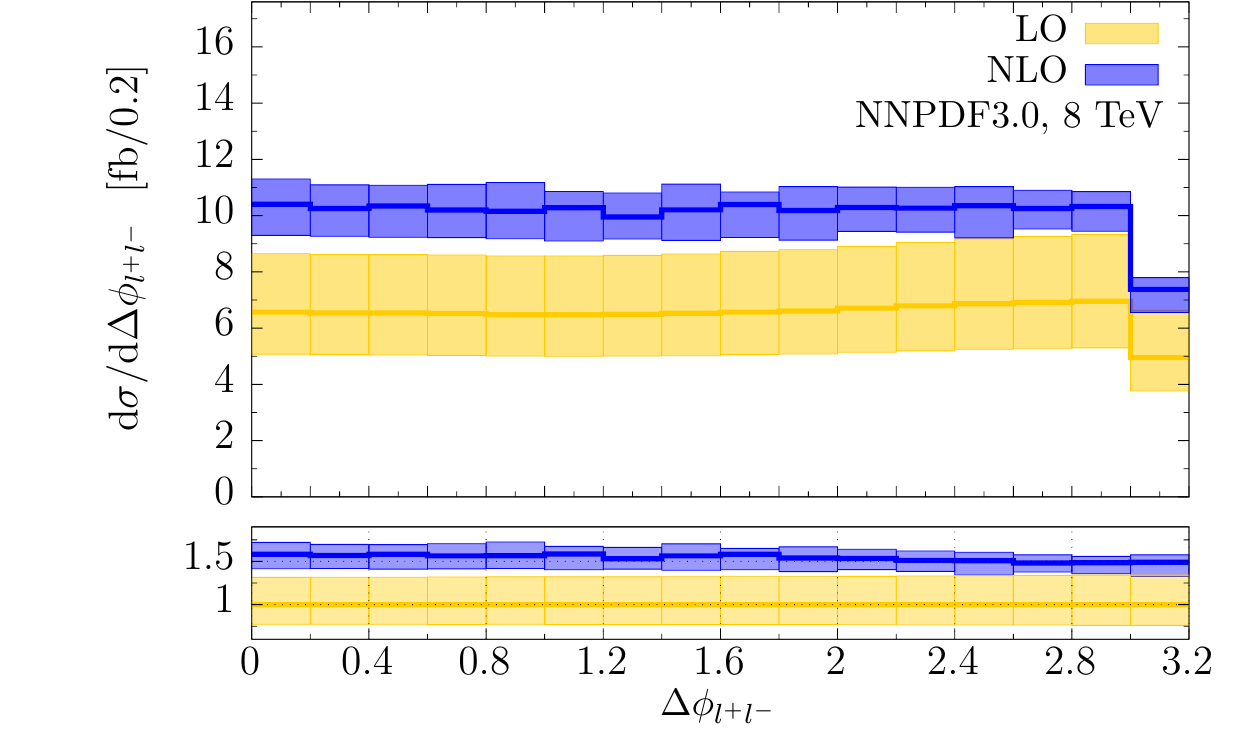}
\includegraphics[scale=0.5]{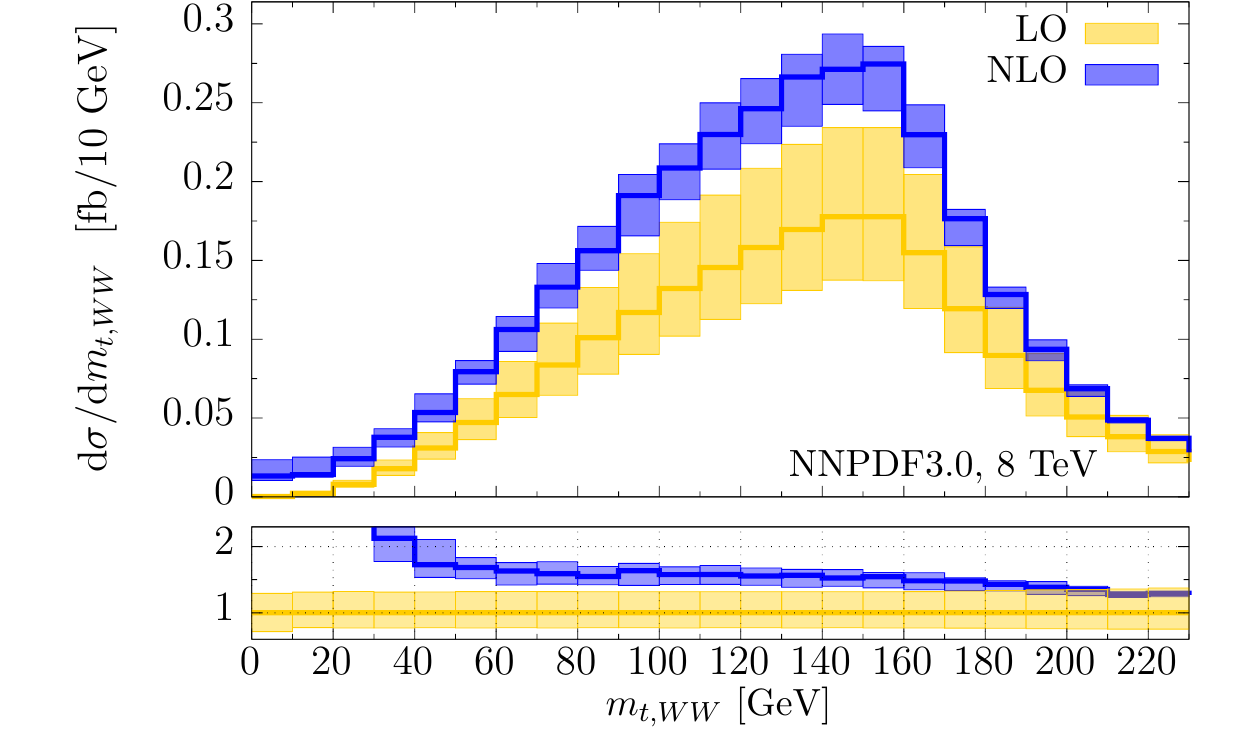}
\caption{\label{fig:distr2} The azimuthal angle between the charged leptons $\Delphi$ (upper plot), 
and the transverse mass of the $\WpWm$ system $\mTWW$ (lower plot), 
 in 
$gg \to \WpWm \to  \nu_e e^+\mu^- \bar{\nu}_{\mu}$ process at the $\sqrt{s}=8$ TeV LHC. LO 
results are shown in yellow, NLO results are shown in blue. 
The central scale is $\mu=m_W$; the scale variation bands correspond to  scale variations by a factor of two in 
either direction. 
The lower panes show the ratios of the LO and NLO distributions at each scale to the LO distribution at the central scale.
}
\end{figure}

We note that gluon fusion contributions both at leading and next-to-leading order are 
less important for $pp \to W^+W^-$ compared to 
 $pp \to ZZ$. Indeed, in the latter case  the corrections to the gluon fusion process were found 
to 
increase the NNLO corrections by approximately 50\% \cite{Caola:2015psa}  and move it beyond the 
estimated uncertainty  of the NNLO result.  The reason gluon fusion is more important for $ZZ$ than for the $W^+W^-$ 
final state is a consequence of the fact that the quark initiated production cross section for 
$pp \to W^+W^-$ and the uncertainties of the final result  are about a factor of seven larger than the quark initiated 
cross section for $pp \to ZZ$,  while the gluon fusion contribution to $W^+W^-$ process 
is only three times larger. 

We repeat the calculation for proton-proton collisions at 
$13~{\rm TeV}$. For the 
$gg \to \WpWm \to \nu_e e^+\mu^- \bar{\nu}_{\mu}$ process, we 
find the LO and the NLO cross sections, 
\beq
\sigma^{W^+W^-}_{gg,\mrm{LO}} = 56.5^{+15.4}_{-11.5}~\mrm{fb},\;\;\;\;\sigma^{W^+W^-}_{gg,\mrm{NLO}} = 79.5^{+4.2}_{-5.9}~\mrm{fb}.
\eeq
The NLO corrections increase the cross section by a factor of $1.2-1.6$, with an increase of 1.4 at 
the central scale. The relative size of QCD radiative corrections is, therefore, similar to that  
at $8~{\rm TeV}$.   The consequences of this increase for the NNLO QCD prediction of $pp \to W^+W^-$ cross sections 
are again similar to what was described earlier for the $8~{\rm TeV}$ case;  the NLO QCD corrections 
to $gg \to W^+W^-$ increase the full NNLO cross section by about $2 \%$ which, roughly, corresponds 
to the scale uncertainty of the   NNLO QCD computation \cite{Gehrmann:2014fva}.

Next, we discuss kinematic distributions. We present results for the 8 TeV LHC. We have 
also studied kinematic distributions at 13 TeV and found a qualitatively similar behavior. 
A representative sample for the $8~{\rm TeV}$ LHC is 
shown in Figs.~\ref{fig:distr1} and~\ref{fig:distr2}. In Fig.~\ref{fig:distr1} we display  the 
positron transverse momentum distribution $\pTlp$ and  the distribution of the 
invariant mass of the dilepton system $\mll$. In Fig.~\ref{fig:distr2} we present the distribution 
of the azimuthal opening angle between the charged leptons $\Delta \phi_{\ell^+\ell^-}$  and the transverse mass of the $\WpWm$ system 
defined as 
\begin{equation}
\mTWW=\sqrt{2p_{\perp,\ell^+\ell^-}E_{\perp,\mrm{miss}}(1-\cos\tilde \phi)}.
\end{equation}
In the definition 
of the transverse mass, we introduced the following notation:
$p_{\perp,\ell^+\ell^-}$ is the transverse momentum of the $\ell^+\ell^-$ system, 
$E_{\perp,\mrm{miss}}$ is the missing energy, 
and $\tilde \phi$ is the azimuthal angle between the direction of the $\ell^+\ell^-$ system and 
the missing momentum.  We observe that for all kinematic distributions shown in Figs.~\ref{fig:distr1} and~\ref{fig:distr2},
with the exception   of the $\mTWW$  one,   the NLO results   can be obtained 
from the LO results by re-scaling the latter by the constant factor determined by the NLO QCD effects 
in the total cross section. 
The situation is different for the 
$\mTWW$ distribution, where the LO distribution vanishes at low values of $\mTWW$, leading to an infinite 
relative correction  in this kinematic regime.
This behavior is easily understood. Indeed, vanishing  of $\mTWW$ requires all  leptons in the final 
state to be  collinear.  This is not possible at LO but may occur  at NLO when 
the $W^+W^-$ system as a whole recoils against an additional jet in the final state.\footnote{
This interpretation is of course independent of the initial 
state; therefore, this effect should be  seen in 
$q\bar{q} \to \WpWm$ at NLO, if no cuts are placed on the leptons. We have checked that this is indeed 
the case using the program MFCM~\cite{mcfm}.} 
We also note that, with the exception of the last bin, the $\Delphi$ distribution 
is remarkably uniform at LO, and this uniformity is maintained at NLO. This is an interesting feature since the
Higgs-mediated process $gg \to H^* \to W^+W^- \to 2l2\nu$ produces a larger number of charged lepton pairs with a 
small relative opening angle $\Delphi$.

We  now turn to the discussion of the fiducial cross sections  defined by a set of cuts used by 
the ATLAS collaboration \cite{ATLAS-CONF-2014-033} for measurements 
with $ee$, $\mu\mu$, and $e\mu+\mu e$ final states.   These cuts are displayed in a concise way 
in Table~1 of Ref.~\cite{Monni:2014zra} and we do not repeat them here. However, we note 
that these cuts include a veto on events with jets with the transverse momentum that exceeds $25~{\rm GeV}$. 
This is an important cut since it reduces the amount of real radiation at NLO and, therefore, is expected 
to reduce  the magnitude of radiative corrections compared to the inclusive cross section case.

\begin{table}[t]
\centering
\def\arraystretch{1.5}
\begin{tabular}{|c|c|c|c|}
\hline
& $\sigma_{\mu\mu,\rm{8~TeV}}$ & $\sigma_{ee,\rm{8~TeV}}$ & $\sigma_{e\mu,\rm{8~TeV}}$ \\
\hline
$\sigma_{gg,\rm LO}$   [fb]    & $5.94^{+1.89}_{-1.35}$ & $5.40^{+1.71}_{-1.23}$ & $9.79^{+3.13}_{-2.24}$ \\
$\sigma_{gg,\rm NLO}$ [fb]     & $7.01^{-0.36}_{-0.17}$ & $6.40^{-0.32}_{-0.16}$ & $11.78^{-0.46}_{-0.34}$ \\
\hline
\end{tabular}
\caption{\label{tab:fidxsec} LO and NLO gluon-initiated 
fiducial cross sections for in the $ee$, $\mu\mu$, and $e\mu$ decay channels. 
The kinematic cuts are defined in Ref.~\cite{ATLAS-CONF-2014-033}. The central 
value corresponds to $\mu = \mu_0$; the upper (lower) value to $\mu =0.5\mu_0~(2\mu_0)$, respectively.
We remind the reader that these numbers do not include contributions from the third generation, see
text for details.
}
\end{table}

In Table~\ref{tab:fidxsec}, we present the fiducial volume cross sections
for the gluon-initiated process at LO and NLO QCD in these three channels.
In order to accurately account for the cuts, these results are computed allowing the $W$-bosons to 
be off  the mass shell.
The NLO QCD values for fiducial  cross sections appear to be maximal at the central scale.
For our choice of the central scale, the NLO corrections increase the fiducial cross sections by $18\%-20\%$, 
independent of the decay channel.
This is substantially smaller than the relative size of radiative corrections 
found for the inclusive cross section. As already mentioned, this large difference 
between corrections to inclusive and fiducial volume cross sections is explained by 
the presence of a jet veto in the ATLAS cuts which 
removes real-emission contributions with a hard gluon. Since the hard gluon radiative 
cross section is positive, the NLO cross section with a jet veto is smaller than the 
cross section without it. A similar effect is known in Higgs production in gluon fusion~\cite{Catani:2001cr}.

Our  observation of smaller radiative corrections in the fiducial volume cross section 
is important since it points towards potential problems  with extrapolating 
fiducial volume cross sections to their inclusive values. In the case of $gg \to VV$ such extrapolations 
completely ignore all the subtleties related to the gluon fusion channel since NLO QCD corrections to 
this mechanism of vector boson production are not included in Monte Carlo event generators.  
Matching our computation to existing NLO parton shower event generators is then desirable.
While this may be challenging technically since the LO process is loop-induced, 
it does not require any conceptual modification 
of existing techniques to combine fixed order computations and parton showers. 

We would like to examine the effects of the NLO corrections to the $gg$ channel shown in 
Table~\ref{tab:fidxsec} on the existing theoretical calculations of the fiducial cross sections.
We compute these fiducial cross sections using MCFM~\cite{mcfm} and the cuts from Ref.~\cite{Monni:2014zra}.
Included in this calculation are the $q\bar{q}$ contributions\footnote{Although we consistently talk about  $q \bar q$ 
contributions, the $q g$ initiated processes are, of course, included, following the standard routine 
of perturbative QCD computations.}  at NLO QCD,  the Higgs production $pp \to H \to W^+W^-$ at NLO QCD 
and the LO $gg$ contributions through quark loops of all flavors, 
with the top mass taken as $m_t=172.5$ GeV and the Higgs signal/background interference at LO QCD. 
We then replace the LO massless $gg$ cross sections in the fiducial volume with the corresponding NLO values.
The $8~{\rm TeV}$ cross sections (in fb) for the $\mu\mu$, $ee$ and $e\mu+\mu e$ decay channels become\footnote{
The NLO $q\bar q$ and LO $gg$ results have opposite scale dependence, so their naive combination would lead
to an accidentally small scale variation uncertainty. If the $gg$ channel is included at NLO, the total
uncertainty is dominated by the $q\bar q$ channel so a precise procedure of how to combine the $q\bar q$ and $gg$ 
uncertainties is not important.}
\beq
\sigma^{q\bar q + H+ gg, \rm NLO}_{\mu\mu,ee,e\mu+\mu e} = (72.0^{+1.3}_{-2.1},\; 66.3^{+1.2}_{-1.7},\; 337.3^{+6.3}_{-4.5}).
\label{eq6}
\eeq

Theoretical results in Eq.(\ref{eq6}) should be compared with results of the ATLAS $8~{\rm TeV}$ measurement 
\beq
\sigma_{\mu\mu,ee,e\mu+\mu e}
= ( 74.4^{+8.1}_{-7.1},\; 68.5^{+9.0}_{-8.0},\; 377.8^{+28.4}_{-25.6}),
\eeq
where we combined statistical, systematic and luminosity uncertainties in quadratures.  We see that 
the electron and muon channels agree perfectly whereas the central value of the $e\mu + \mu e$ channel
differs by about 1.5 standard deviations.  However, this picture is somewhat misleading, 
since we have not included the NNLO QCD 
corrections to the $q \bar q $ channel in the theory predictions in Eq.(\ref{eq6}).
While these corrections are unknown in the fiducial region, it is perhaps interesting to see
what happens if one estimates them  by re-scaling NNLO QCD corrections 
to the inclusive cross section by the ratio of 
fiducial and inclusive cross sections. In this case we find that the missing NNLO QCD corrections can increase the 
cross sections in Eq.(\ref{eq6}) by ${\cal O}(4 - 20)~{\rm fb}$ for $ee(\mu\mu)$ and $e\mu+ \mu e$ channels, respectively. 
Such an increase would make the theory prediction and experimental results agree to within one standard deviation 
 for each 
of the three channels. 
 
{\bf In summary}  We have calculated the NLO QCD corrections to the $gg \to \WpWm \to l^+_1 \nu_1 l^-_2 \bar \nu_2$ process at the LHC.
These corrections increase the gluon fusion cross section 
by $20\%-80\%$, depending on the center-of-mass energy and the scale choice. The impact of these 
corrections on the $pp \to W^+W^-$ production cross section is moderate; they increase the NNLO QCD theory 
prediction by about two percent, which is comparable to the current estimate of the theoretical uncertainty at NNLO.
We have also calculated the $gg \to W^+W^-$ cross section through NLO in perturbative QCD 
subject to kinematic 
cuts used by the ATLAS collaboration 
to measure the $pp \to W^+W^-$  cross section.  For the fiducial cross section, 
we found  a smaller increase of around $20\%$ for our central scale choice. Nevertheless, 
this contribution further increases the fiducial volume cross section, moving the theoretical result 
closer to the experimental one.

{\bf Acknowledgments}
We are grateful to S. Pozzorini and, especially, to J. Lindert for their help in checking the $gg \to \WpWm+g$ 
scattering amplitude computed in this paper against the implementation in OpenLoops~\cite{Cascioli:2011va,openloops}.
We would like to thank P. Monni and G. Zanderighi for discussions on the fiducial measurements and
for providing their implementation of the ATLAS experimental cuts. 
The research reported in this paper is partially supported by 
BMBF grant 05H15VKCCA. 
F.C. and K.M. thank the Mainz Institute for Theoretical Physics (MITP) for
hospitality and partial support during the program \emph{Higher Orders and Jets for LHC}.

\end{document}